\documentclass[11pt,english,twoside]{article}

\usepackage[T1]{fontenc}
\usepackage[latin1]{inputenc}
\usepackage[english]{babel}
\usepackage{lmodern}
\usepackage{a4wide}
\usepackage{amssymb, amsmath, amsthm}
\usepackage{slashed}
\usepackage{float}
\usepackage{graphicx}
\usepackage[dvips]{epsfig}
\usepackage{psfrag}
\usepackage{lscape}
\usepackage[all]{xy}
\usepackage{hyperref}
\usepackage{enumerate}
\usepackage{dsfont}

\usepackage{mathabx}

\newcommand{\beq}{\begin{equation}}
\newcommand{\eeq}{\end{equation}}
\def\bea#1\eea{\begin{align}#1\end{align}} \newcommand{\nn}{\nonumber}

\def\del {\partial}

\def\eee{{\cal E}}
\def\hhh{{\cal H}}

\def\R {\mathcal{R}}
\def\L {\mathcal{L}}

\def\tL{\tilde{{\cal L}}}
\def\te{\tilde{e}}

\def\tg{\tilde{g}}

\def\tp {\tilde{\phi}}
\def\b {\beta}
\def\p {\phi}
\def\tx{\tilde{x}}

\def\LD {{\cal L}_{{\rm DFT}}}

\renewcommand{\i}{\ensuremath{\textnormal{i}}}

\makeatletter
\@addtoreset{equation}{section}
\makeatother

\begin{document}

% ---------------------------------------- Title page
\begin{titlepage}

\rightline{\small MPP-2013-45}

\vskip 2.4cm

{\fontsize{18.2}{21}\selectfont
  \flushleft{\noindent\textbf{Non-geometric fluxes versus (non)-geometry}} }

\vskip 0.2cm
\noindent\rule[1ex]{\textwidth}{1pt}
\vskip 0.8cm

\noindent\textbf{David Andriot$^{a,b}$}

\vskip 0.6cm
\begin{enumerate}[$^a$]
\item \textit{Max-Planck-Institut f\"ur Physik\\F\"ohringer Ring 6, 80805 M\"unchen, Germany}
\vskip 0.2cm
\item \textit{Arnold-Sommerfeld-Center for Theoretical Physics\\Fakult\"at f\"ur Physik, Ludwig-Maximilians-Universit\"at M\"unchen\\Theresienstra\ss e 37, 80333 M\"unchen, Germany}
\end{enumerate}

\noindent {\small{\texttt{andriot@mpp.mpg.de}}}

\vskip 3.1cm

\noindent This paper is based on a talk given at String-Math 2012 in Bonn, Germany, and contributes to the proceedings of this conference.\\

\begin{center}
{\bf Abstract}
\end{center}

\noindent Non-geometry has been introduced when considering a new type of string backgrounds, for which stringy symmetries serve as transition functions between patches of the target space. Then, some terms in the potential of four-dimensional gauged supergravities, generated by so-called non-geometric fluxes, have been argued to find a higher-dimensional origin in these backgrounds, even if a standard compactification on those cannot be made. We present here recent results clarifying the relation between these two settings. Thanks to a field redefinition, we reformulate the NSNS Lagrangian in such a way that the non-geometric fluxes appear in ten dimensions. In addition, if an NSNS field configuration is non-geometric, its reformulation in terms of the new fields can restore a standard geometry. A dimensional reduction is then possible, and leads to the non-geometric terms in the four-dimensional potential. Reformulating similarly doubled field theory, we get a better understanding of the role of the non-geometric fluxes, and rewrite the Lagrangian in a manifestly diffeomorphism-covariant manner. We finally discuss the relevance of the field redefinition and the non-geometric fluxes when studying the non-commutativity of string coordinates.

\vfill

\end{titlepage}

% ---------------------------------------- Table of contents
\tableofcontents

% ---------------------------------------- Main text
%%%%%%%%%%%%%%%%%%%%%%%%%%%%
\section{Introduction}\label{sec:intro}

String backgrounds with non-trivial fluxes on an internal space are crucial for phenomenology. Solutions of four-dimensional supergravity with non-geometric fluxes are in that respect rather promising. Some of them are indeed among the few examples of metastable de Sitter solutions \cite{dS}, and others allow to achieve a full moduli stabilisation \cite{stab}. Unfortunately, the uplift to ten dimensions of such solutions has been so far rather unclear, as it should involve a ten-dimensional non-geometry. One aim of the results \cite{Andriot:2011uh, Andriot:2012wx, Andriot:2012an, Andriot:2012vb} presented here is to understand better the relation between these two different settings. Before going any further, let us give a brief account on these ten- and four-dimensional perspectives. We restrict here the discussion to supergravity, and only consider the NSNS sector.

\begin{itemize}
\item In ten dimensions

We first consider the target space of a string, divided locally into patches. If some fields are living on each of these patches, they can be defined globally by ``gluing'' them from one patch to the other using transition functions. The latter, for a standard differential geometry, are the diffeomorphisms (completed with gauge transformations), i.e. the usual symmetries of a point-like field theory. String theory has actually more symmetries, for instance T-duality on some backgrounds. The essential idea of non-geometry is then to use these more stringy symmetries to glue the fields from one patch to the other \cite{Hellerman:2002ax, Dabholkar:2002sy, Flournoy:2004vn}. From the string theory point of view, the resulting geometry is equally fine, and could serve as a background. However, such stringy transition functions take us away from standard differential geometry, hence the name of ``non-geometry''. From the point of view of an effective theory in the target space, this situation can be more problematic. In practice, the ten-dimensional supergravity fields would look ill-defined, as they would not be single-valued in the usual sense. One typically faces global issues with such field configurations.

\item In four dimensions

Some four-dimensional gauged supergravities have in their super- or scalar potential specific terms generated by the so-called non-geometric fluxes. The latter are quantized objects $Q_p{}^{mn}$ and $R^{mnp}$, and can be identified with some structure constants of the gauge algebra; in other words they correspond to specific gaugings of the four-dimensional supergravity \cite{Shelton:2005cf, Dabholkar:2002sy, Dabholkar:2005ve}. As discussed above, these non-geometric terms of a four-dimensional potential can be of phenomenological interest.

\end{itemize}

One reason to make a link between these two settings can be found in the study of a simple non-geometric field configuration \cite{torex} that we call the toroidal example. It has the property of being related to geometric configurations when applying standard T-duality transformations along isometries. Geometric backgrounds lead after compactification to specific terms in the four-dimensional potential, such as those generated by the NSNS $H$-flux or by the so-called ``geometric flux'' (related to the curvature of the internal manifold). One can then determine what is the corresponding transformation of these type of terms when performing a T-duality at the ten-dimensional level. The result is that they can transform into the terms generated by $Q$ or $R$ (these rules are described by the so-called T-duality chain). It is in particular the case for the toroidal example, where its non-geometric configuration should then correspond to a four-dimensional $Q$-flux (hence a ``non-geometric'' flux). However, this correspondence is not really constructive, as there is no direct determination of $Q$ given the non-geometry. For more details on this discussion, and for a review on non-geometry, we refer the reader to \cite{Andriot:2011uh}.

A typical relation between a ten- and a four-dimensional supergravity would be a compactification, so can one get the non-geometric terms of the potential from such a process? To start with, the specific dependence of these terms in the scalar fields cannot be reproduced from a standard ten-dimensional supergravity. In addition, the index structure of $Q$ and $R$ is very special, and can a priori not be found in a flux or a field in ten dimensions; their origin is thus unclear. Finally, we mentioned that ten-dimensional non-geometric field configurations have been argued to be related to four-dimensional non-geometric potential terms. It is therefore tempting to consider a compactification on those. However, the global issues, characteristic of these field configurations, make a standard dimensional reduction not possible; in particular one cannot integrate the fields properly.\\

Here, we present recent progress \cite{Andriot:2011uh, Andriot:2012wx, Andriot:2012an} in relating ten-dimensional non-geometry and four-dimensional non-geometric fluxes. This is made possible thanks to a reformulation of standard supergravity. More precisely, we consider a field redefinition to be performed on the NSNS fields. The NSNS Lagrangian can this way be rewritten into a new ten-dimensional Lagrangian, in which the non-geometric $Q$- and $R$-fluxes appear. In addition, starting with a standard non-geometric field configuration, the redefined fields and new Lagrangian can be globally well-defined. One can then perform the dimensional reduction, and doing so gives precisely the non-geometric terms in the scalar potential. This way, the ten-dimensional and four-dimensional perspectives are finally related.

While these ideas are presented in section \ref{sec:reformsugra}, we discuss in section \ref{sec:broader} the interesting roles played by the field redefinition and the non-geometric fluxes in broader contexts. In double field theory (DFT) \cite{DFT}, they help to reformulate the DFT Lagrangian in a manifestly diffeomorphism-covariant manner. We then get a better understanding of the non-geometric fluxes: the $R$-flux is a tensor, while the $Q$-flux serves more as a connection; this is analogous to the NSNS $H$-flux, and the geometric flux. Another topic is the non-commutativity of string coordinates, that we studied in \cite{Andriot:2012vb}: interesting relations occur between non-geometry, non-geometric fluxes and non-commutativity. We come back to them in more details.\\

\noindent {\bf Note}: Due to huge delays in the publication process, some results presented in the initial version of this paper have been a little outdated. The necessary updates are now provided in three addenda at the end of the paper, to which we refer in the main text.

\section{Reformulation of the NSNS sector of supergravity}\label{sec:reformsugra}

\subsection{The field redefinition}\label{sec:fieldredef}

An important object in the reformulation is the field $\b^{mn}$, which is an antisymmetric bivector. Our first motivations to consider this field came from \cite{Grange:2006es, Grange:2007bp, Grana:2008yw}, where generalized complex geometry tools are used in supergravity to study ten-dimensional non-geometry. Several arguments are then put forward to indicate that $\b$ is a good object to characterise the presence of non-geometry. Relations between $\b$ and the non-geometric $Q$- and $R$-fluxes are even proposed (see \cite{Andriot:2011uh} for a more detailed account on these ideas). We concluded that making $\b$ appear could lead to a reformulation of ten-dimensional supergravity which would provide an origin to the four-dimensional non-geometric fluxes. A way of introducing $\b$ is by considering a different generalized vielbein than the standard NSNS one \cite{Grana:2008yw}. The idea used in \cite{Andriot:2011uh} is then that $\b$ would appear through a reparameterization of the generalized metric, associated to the choice of this different generalized vielbein. More precisely, the generalized metric $\hhh$ usually depends on the standard NSNS metric $g_{mn}$ and Kalb-Ramond field $b_{mn}$, and we considered the following reparametrization
\beq
\hhh= \begin{pmatrix} g-b g^{-1}b & b g^{-1} \\ -g^{-1} b & g^{-1}\end{pmatrix}
= \begin{pmatrix} \tg & -\tg\b \\ \b \tg & \tg^{-1}-\b \tg \b \end{pmatrix} \ ,
\eeq
where $\tg_{mn}$ is a new metric. The above can be rewritten in several manners, among which
\beq
\begin{array}{l|}
g= (\tg^{-1}+ \b)^{-1} \tg^{-1} (\tg^{-1}- \b)^{-1} \\
b=-(\tg^{-1}+ \b)^{-1} \b (\tg^{-1}- \b)^{-1}
\end{array}
\quad \Leftrightarrow \quad (g + b) = (\tg^{-1}+ \b)^{-1} \ ,\label{fieldredef}
\eeq
where the last equality is useful in DFT (see section \ref{sec:DFTrewrite}). The standard NSNS dilaton $\p$ also needs a counterpart, so we introduced the new dilaton $\tp$, defined as follows
\beq
e^{-2\tp} \sqrt{|\tg|}=e^{-2\phi} \sqrt{|g|} \ ,
\eeq
so that the supergravity measure gets preserved.

To summarize, we first have a field redefinition from the standard NSNS fields $(g,b,\p)$ to the new set of fields $(\tg,\b,\tp)$. A natural question is then that of the fate of the standard NSNS Lagrangian under this redefinition. As $\b$ has been proposed to be related to the non-geometric fluxes, they could appear in the resulting ten-dimensional Lagrangian. This is indeed the result obtained in \cite{Andriot:2011uh, Andriot:2012an}, that we now detail. For completeness, let us indicate the existence of an alternative field redefinition, proposed in \cite{ralph}.

\subsection{Rewriting the NSNS Lagrangian}\label{sec:rewrit}

The standard NSNS Lagrangian is given by
\beq
\L = e^{-2\phi} \sqrt{|g|} \left(\R(g) + 4 (\del \phi)^2 - \frac{1}{12} H_{mnp} H_{qrs}\ g^{mq} g^{nr} g^{ps} \right) \ , \label{L}
\eeq
where $\R(g)$ is the Ricci scalar associated to the metric $g$ for a Levi-Civita connection, the $H$-flux is given by $H_{mnp}=3\ \del_{[m}b_{np]}$, and we refer to \cite{Andriot:2012an} for more conventions. Performing the field redefinition directly in $\L$ is a rather tedious computation. For instance, the Ricci scalar $\R(g)$ is given by the one, $\R(\tg)$, associated to $\tg$, together with eleven lines of other terms (see equation (B.15) in \cite{Andriot:2012an}). It is therefore remarkable that most of these terms get in the end cancelled by others, coming from the two remaining terms of \eqref{L} once they are also rewritten. This leaves us with, in comparison, a surprisingly simple resulting Lagrangian $\tL$, given by
\bea
\tL& = \L - \del (..) \label{tL}\\
& = e^{-2\tp} \sqrt{|\tg|} \left(\R(\tg) + 4(\del \tp)^2 - \frac{1}{4} Q_p{}^{mn} Q_q{}^{rs}\ \tg^{pq} \tg_{mr} \tg_{ns} + ... - \frac{1}{12} R^{mnp} R^{qrs}\ \tg_{mq} \tg_{nr} \tg_{ps} \right) \nn
\eea
where we introduce ten-dimensional $Q$- and $R$-fluxes as
\beq
Q_p{}^{mn}=\del_p \b^{mn} \ , \ R^{mnp}=3\ \b^{q[m}\del_q \b^{np]} \ .\label{10dnongeoflux}
\eeq
Another, more efficient, method to derive this result is to use DFT as a tool \cite{Andriot:2012wx, Andriot:2012an}. We give more details about it in section \ref{sec:DFTrewrite}. Note as well that we first computed $\tL$ in \cite{Andriot:2011uh} using a simplifying assumption, given by $\b^{mn}\del_n \cdot =0$ where the $\cdot$ is a placeholder for any field. In that case, the dots in \eqref{tL} and the $R$-flux term vanish. Without this assumption, the terms in the dots should be thought of as going together with the $Q$-flux term (see \cite{Andriot:2012an}), leading to a possible redefinition of the actual $Q$-flux (see addendum 1).

Applying the field redefinition to the NSNS Lagrangian thus gives, up to a total derivative, a new ten-dimensional Lagrangian $\tL$. The latter depends on new objects denoted $Q$ and $R$, which have the same index structure as their four-dimensional counterparts. Is there a more precise relation to the four-dimensional non-geometric fluxes? To verify this, one should perform a dimensional reduction on $\tL$. This was done in \cite{Andriot:2011uh, Andriot:2012an} by studying the dependence of the four-dimensional scalar potential on two moduli: the volume and the four-dimensional dilaton. This is enough to see that the dimensional reduction of $\tL$ gives the expected non-geometric terms of the potential (defined in \cite{Hertzberg:2007wc}), while the standard NSNS Lagrangian $\L$ does not lead to such terms. We conclude that for the first time, these terms get this way a ten-dimensional supergravity origin. In addition, we know that the ten- and four-dimensional non-geometric fluxes are related, but the precise relation remains to be established (see addendum 1).

An important aspect of the dimensional reduction is the global behaviour of the background. The reformulation we have done so far is somehow made formally. In addition, the quantities involved (fields and Lagrangian) are local. But when performing a dimensional reduction, one needs to consider the action, and integrate over the background fields. This implies to consider the global behaviour of an actual background configuration of fields. This is where the discussion on non-geometry becomes crucial.

\subsection{The global aspects}\label{sec:global}

For a geometric background, the fields are by definition globally well-defined, and so is the NSNS Lagrangian $\L$. One can therefore perform a dimensional reduction on it, and this gives the standard terms due to the $H$-flux and the internal curvature. Let us now consider a non-geometric NSNS field configuration: as explained already, the global issues of the fields, and consequently of $\L$, make the integration of the Lagrangian over some internal directions not possible. It is therefore unclear how to derive a four-dimensional potential from such a background. What we have shown though in \eqref{tL} is that $\L$ can be rewritten as
\beq
\L(g,b,\p)= \tL(\tg, \b, \tp) + \del(\dots) \ . \label{LtL}
\eeq
A nice feature of this reformulation is that despite the ill-definedness of $\L$, the new fields and new Lagrangian $\tL$ can be globally well-defined. This is precisely what happens for the toroidal example mentioned in the introduction, and we expect it to hold for other examples as well \cite{Andriot:2011uh} (see addendum 3). Given the equality \eqref{LtL}, such a situation can only occur if the total derivative $\del(\dots)$ is also ill-defined. For the toroidal example, it is not single-valued and therefore does not integrate to zero.

In that case, we propose to consider $\tL$ as the correct effective description of string theory. In other words, for string theory on a non-geometric background, one should not consider the NSNS {\it action} to start with, as it does not make sense, but rather the action given by a well-defined $\tL$ whenever it exists. The dimensional reduction of $\tL$ is then allowed, and as we explained above, it gives the non-geometric terms of the four-dimensional potential. Let us emphasise once more that the global aspects are the whole subtlety of the discussion: one would naively think that a field redefinition does not change much; but on a non-geometric configuration, it can restore a standard geometry, at the cost of introducing new fluxes, and then leads a previously unreached four-dimensional theory.

Thanks to this procedure, we establish a relation between the ten-dimensional non-geometry and the four-dimensional non-geometric fluxes: given a non-geometric field configuration, first perform a field redefinition to a well-defined set of fields, and then dimensionally reduce to get the four-dimensional non-geometric terms of the potential. For the toroidal example, this procedure works perfectly, and the ten-dimensional $Q$-flux \eqref{10dnongeoflux} matches its four-dimensional counterpart, given by the T-duality chain. We discuss in \cite{Andriot:2011uh, Andriot:2012an} the extension of such a situation to other examples (see addendum 3).

\section{The non-geometric fluxes in broader contexts}\label{sec:broader}

\subsection{Reformulation of double field theory}

\subsubsection{First properties and field redefinition}\label{sec:DFTrewrite}

Double field theory \cite{DFT} is defined on a doubled space, where to the standard coordinates $x^m$ are added so-called dual coordinates $\tilde{x}_m$; the resulting space has twice the usual number $d$ of dimensions. Coordinates are gathered into $X^M=(\tilde{x}_m, x^m)$ and one introduces the associated derivative $\del_M=(\tilde{\del}^m,\del_m)$. All the fields of DFT depend at first on both sets of coordinates. We consider here only the NSNS sector; one formulation of DFT is then given in terms of the combination $\eee_{mn} (x,\tilde{x}) = (g+b)_{mn} (x,\tilde{x})$ and the dilaton.

A first property of DFT is its $O(d,d)$ invariance, where transformations are encoded as
\beq
{\rm For}\ h= \begin{pmatrix} A & B \\ C & D \end{pmatrix} \in O(d,d)\ , \ X' = h X\ ,\ {\cal E}^{\prime}(X^{\prime}) = (A{\cal E}(X)+B)(C{\cal E}(X)+D)^{-1} \ .
\eeq
These reproduce the standard fractional linear transformation of T-duality, but also go beyond it. Indeed, acting along non-isometry directions is allowed here, when transforming in particular the coordinates, so this $O(d,d)$ action is more general than T-duality.

Another property is the requirement of the strong constraint, applied on all fields and their products. A consequence is that the fields depend locally only on half of the coordinates in $X^M$. We consider in the following this constraint in the form of $\tilde{\del}^m =0$, i.e. by loosing any dependence in the dual coordinates. Applying this to the DFT Lagrangian $\LD$ gives the standard NSNS Lagrangian up to a total derivative
\beq
\LD (\eee, \p) \stackrel{\tilde{\del}^m=0}{====} {\cal L}(g,b,\phi) +\del(\dots) \ .\label{LDFT1}
\eeq

These two properties are used to obtain the equality of Lagrangians \eqref{tL}. To start with, performing the field redefinition within DFT is simple, as it essentially involves the quantity $\eee$. Indeed, as indicated above, the field redefinition \eqref{fieldredef} can take the form $\eee=(g+b)=(\tg^{-1}+\b)^{-1}$. The last inverse power makes things a little more involved though, and one uses in addition the $O(d,d)$ invariance to get the result (we refer the reader to \cite{Andriot:2012an} for the details of the procedure). We finally obtain the DFT Lagrangian expressed in terms of the new fields. Using again the strong constraint, the resulting Lagrangian is precisely $\tL$ given in \eqref{tL} up to a total derivative
\beq
\LD(\tg,\b,\tp) \stackrel{\tilde{\del}^m=0}{====} \tL(\tg,\b,\tp) +\del(\dots) \ .\label{LDFT2}
\eeq
Thanks to \eqref{LDFT1} and \eqref{LDFT2}, we recover the equality \eqref{tL} relating the standard NSNS Lagrangian $\L$ to the new Lagrangian $\tL$.

\subsubsection{Diffeomorphism covariance}\label{sec:diffeocov}

Another property of DFT is its gauge invariance under double diffeomorphisms. Those are the transformations of the two sets of coordinates that we parameterize as $x^m\rightarrow x^m-\xi^m(x ,\tx)$, $\tilde{x}_m\rightarrow \tilde{x}_m-\tilde{\xi}_m (x,\tilde{x})$. The DFT Lagrangian is in most of its formulations not manifestly covariant under these transformations. Let us illustrate this point with $\LD$ depending on the new fields, which can be expressed as
\beq
\!\!\!\!\! \LD(\tg,\b,\tp ) = e^{-2\tp} \sqrt{|\tg|} \left( \R (\tg,\partial)  + 4(\partial\tp)^2  + \R (\tg^{-1},\tilde{\partial})  + 4 (\tilde{\partial}\tp)^2 -\frac{1}{2} Q^2 -\frac{1}{2} R^2 +\dots \right) \label{LDFT3}
\eeq
where $\R (\tg,\partial)$ is the standard Ricci scalar already mentioned, and $\R (\tg^{-1},\tilde{\partial})$ has exactly the same expression with all up and down indices exchanged. In addition, the $Q^2$ and $R^2$ denote the same expressions as in \eqref{tL} with indices contracted by metrics. The DFT $R$-flux has however an additional term with respect to \eqref{10dnongeoflux}, as first proposed in \cite{Aldazabal:2011nj}
\beq
R^{mnp}= 3\big( \tilde{\partial}^{[m}\b^{np]}  +  \b^{q[m}\partial_{q}\beta^{np]} \big) \ .\label{DFTRf}
\eeq

Under the standard diffeomorphisms $x^m\rightarrow x^m-\xi^m(x )$, resp. the dual diffeomorphisms $\tilde{x}_m\rightarrow \tilde{x}_m-\tilde{\xi}_m (\tilde{x})$, the terms $\R (\tg,\partial)  + 4(\partial\tp)^2$, resp. $\R (\tg^{-1},\tilde{\partial})  + 4 (\tilde{\partial}\tp)^2$, transform covariantly. This behaviour is however lost if one considers the dependence in the other coordinate, and looks at the transformation under the ``unnatural'' diffeomorphism.  For instance, the terms $\R (\tg^{-1},\tilde{\partial})  + 4 (\tilde{\partial}\tp)^2$ a priori do not transform covariantly under the $\xi$-diffeomorphisms $x^m\rightarrow x^m-\xi^m(x , \tx)$. So we constructed in \cite{Andriot:2012wx, Andriot:2012an} a covariant derivative $\tilde{\nabla}^m$ completing $\tilde{\del}^m$ in order to restore covariance under the ``unnatural'' $\xi$-diffeomorphisms. The non-geometric fluxes were an inspiration for the structures involved: indeed, while the $Q$-flux, defined as in \eqref{10dnongeoflux}, is not a tensor, the DFT $R$-flux \eqref{DFTRf} is a tensor under the $\xi$-diffeomorphisms. The covariant derivative $\tilde{\nabla}^m$ is defined by the following action on a vector and a co-vector
\beq
\tilde{\nabla}^{m}V^{n} \ = \ \tilde{D}^{m}V^{n}-\widecheck{\Gamma}_{p}{}^{mn}V^{p} \ , \ \tilde{\nabla}^{m}V_{n} \ = \ \tilde{D}^{m}V_{n}+\widecheck{\Gamma}_{n}{}^{mp}V_{p} \ ,
\eeq
where its two building blocks are given by
\bea
\tilde{D}^{m}  &=   \tilde{\partial}^m + \b^{pm}\partial_{p} \ ,\\
\widecheck{\Gamma}_{p}{}^{mn} &=\frac{1}{2}\tilde{g}_{pq}\left(\tilde{D}^m\tilde{g}^{nq}+  \tilde{D}^n\tilde{g}^{mq}-\tilde{D}^q\tilde{g}^{mn}\right) +\tilde{g}_{pq}\tilde{g}^{r(m}Q_{r}{}^{n)q}-\frac{1}{2}Q_{p}{}^{mn} \ .\nn
\eea
The derivative $\tilde{D}^m$ is tied to the $R$-flux \eqref{DFTRf}, since $R^{mnp} =3\tilde{D}^{[m}\beta^{np]}$. The $Q$-flux also enters these buildings blocks, as providing the non-standard part of the connection $\widecheck{\Gamma}_{p}{}^{mn}$ (the first part of this connection is very analogous to the Levi-Civita connection); see addendum 2.

From these objects, we constructed a Riemann tensor $\widecheck{\R}^{mn}{}_{p}{}^{q}$, a Ricci tensor $\widecheck{\R}^{mn}$ and finally a Ricci scalar $\widecheck{\R}$. We used the latter to rewrite the DFT Lagrangian as
\beq
\LD (\tg,\b,\tp ) = e^{-2\tp} \sqrt{|\tg|} \left( \R (\tilde g,\partial) + 4 (\partial \tp)^2 + \widecheck{\R}(\tilde{D},\widecheck{\Gamma})  + 4\big(\tilde{D}^m\tp +\widecheck{\Gamma}_{p}{}^{pm}\big)^2 -\frac{1}{2} R^2 \right) + \del(\dots) \label{LDFT4}
\eeq
where each term is, at last, manifestly covariant under the $\xi$-diffeomorphisms.

We get a better understanding of the role of the non-geometric fluxes through \eqref{LDFT4}. The $R$-flux always appears as a tensor, and is then analogous to the standard NSNS $H$-flux. The $Q$-flux on the contrary is not a tensor, but rather plays the role of a connection, analogously to the geometric flux. In addition, all the $Q$-flux terms in \eqref{LDFT3}, and in some sense, in \eqref{tL}, are absorbed within the connection $\widecheck{\Gamma}_{p}{}^{mn}$, essentially appearing through $\widecheck{\R}$. This provides some structure to these terms (see addendum 2).

\subsection{Non-commutativity of the string coordinates}

Non-commutativity in string theory first appeared in the context of the open string. By considering open strings ending on a D-brane with a constant $b$-field, their coordinates were found not to commute, as given by $[x^m(\tau),x^n(\tau)]=\i \theta^{mn}$. The non-commutativity parameter $\theta^{mn}$ was defined \cite{Seiberg:1999vs}, together with the so-called open string metric $G_{mn}$, by
\beq
\!\!\!\! G^{mn}=\left(\frac{1}{g+2\pi\alpha' b}\ g\ \frac{1}{g-2\pi\alpha' b} \right)^{mn} \ ,\  \theta^{mn} =-(2\pi\alpha')^2\left(\frac{1}{g+2\pi\alpha' b}\ b\ \frac{1}{g-2\pi\alpha' b} \right)^{mn} \ .
\eeq
As noticed in \cite{Andriot:2012an} and other references therein, it is striking to compare these formulas with the field redefinition \eqref{fieldredef}, that we repeat here in a slightly rewritten form
\beq
\tg^{-1}= (g+ b)^{-1}\ g\  (g - b)^{-1} \ ,\ \b= -(g+ b)^{-1}\ b\ (g- b)^{-1}  \ .
\eeq
Up to conventions on $\alpha'$, they perfectly match. We are however working in a different context, involving closed strings and non-geometry. It would nevertheless be interesting to investigate further this relation.

Non-commutativity for closed strings has been studied more recently. In the serie of papers \cite{Lust:2010iy, Condeescu:2012sp, Andriot:2012vb}, a closed string, on various examples of non-geometric backgrounds, was shown to have some coordinates not commuting: $[{\cal X}^m(\tau,\sigma),{\cal X}^n(\tau,\sigma)]\neq 0 $. Drawing the analogy with the open string, one would think of $\b$ as the non-commutativity parameter. However, in the examples studied, the non-vanishing commutator was rather parameterized by the $Q$-flux times a winding number $N^p$
\beq
[{\cal X}^m(\tau,\sigma),{\cal X}^n(\tau,\sigma)] \sim \i N^p \ Q_p{}^{mn} \ .\label{NC}
\eeq
Before we comment on this result, let us indicate that the $R$-flux as well served as a parameter, in the context of non-associativity \cite{NA}. More precisely, it parameterizes a non-vanishing jacobiator of closed string coordinates. So the new fields introduced, in particular $\b$, and the non-geometric fluxes seem to play a role in characterising non-commutative and non-associative structures in string theory.

The right-hand side of \eqref{NC} was discussed at length in \cite{Andriot:2012vb}, where the non-geometric background considered is that of the toroidal example. For this background, the space is given locally by a two-torus fibered over a base circle. The non-geometry occurs when going around the base: the fiber then requires a T-duality as a transition function. The $Q$-flux characterising this non-geometry, in the sense of section \ref{sec:global}, is simply given by \eqref{10dnongeoflux} and is constant. We discuss in \cite{Andriot:2012vb} several arguments explaining why the non-commutativity of the fiber coordinates is related to the non-geometry; one of them being that the fiber is somehow fuzzy, so a precise determination of the position (in the point particle sense) is not possible. This relation justifies why the non-geometric $Q$-flux parameterizes the non-vanishing commutator. In addition, one can understand for this background the presence in \eqref{NC} of the string winding number along the base circle. Such a wound closed string probes directly the non-geometry of the fiber, as it goes through its non-trivial monodromy when wrapping the base circle. On the contrary, a non-wound string would only probe the local geometry. Thanks to these arguments, the formula \eqref{NC}, or the more precise one given in \cite{Andriot:2012vb}, may be as well valid for a closed string on another non-geometric background of a similar type (those which can be viewed as a T-fold \cite{Hull:2004in}).

\section{Conclusion and outlook}\label{sec:ccl}

We have presented a reformulation of the NSNS sector of supergravity in terms of new fields and fluxes. This reformulation is made through a field redefinition from the standard $(g_{mn},b_{mn},\p)$ to new fields $(\tg_{mn}, \b^{mn}, \tp)$. Rewriting the standard NSNS Lagrangian $\L$ in terms of the new fields gives, up to a total derivative, the new Lagrangian $\tL$. The latter depends on new objects that we identify as ten-dimensional non-geometric fluxes, namely the $Q$-flux given by $Q_p{}^{mn}= \del_p \b^{mn}$ (at least for $\b^{mn}\del_n \cdot =0$) and the $R$-flux $R^{mnp}=3\ \b^{q[m}\del_q \b^{np]}$.

The new ten-dimensional Lagrangian $\tL$ is proposed to provide an uplift to some four-dimensional gauged supergravities, because its dimensional reduction leads generically to non-geometric terms in the scalar potential. A concrete compactification can though only be done on a given background, and the global aspects of the latter are then crucial. For the non-geometric field configuration of the toroidal example, the global issues are cured when using the new fields; a standard differential geometry is also restored. $\tL$ can then be dimensionally reduced in the usual way, while it was not possible with the standard NSNS Lagrangian. For this example, ten-dimensional and four-dimensional $Q$-fluxes match. This procedure establishes a relation between a ten-dimensional non-geometry, and four-dimensional non-geometric fluxes.

Double field theory can also be reformulated in terms of the new fields. While the resulting Lagrangian is not manifestly diffeomorphism-covariant, the non-geometric fluxes exhibit inspiring structures to rewrite it in such a manner. Doing so, the $R$-flux appears to always behave as a tensor, while the $Q$-flux does not, but rather plays the role of a connection.

The new fields and the non-geometric fluxes also appear as relevant quantities to study non-commutativity of string coordinates. In particular, for a closed string on a non-geometric background, the $Q$-flux, together with a winding number, are argued to be the parameters of the non-vanishing commutator.\\

It would be interesting develop more the reformulation of ten-dimensional supergravity. To do so, one should first get a better handle on the $Q$-flux terms that appear in $\tL$, and clarify the relation to the four-dimensional $Q$-flux (see addenda 1 and 2). Then, one could consider extensions to other sectors, such as the Ramond-Ramond sector, where the fluxes are known to have non-geometric counterparts \cite{RR}, or to the gauge fields of heterotic string. For the latter, introducing a generalized metric as in \cite{Andriot:2011iw} should be helpful. Finally, there should as well exist non-geometric counterparts to the D-brane and O-plane sources, possibly related to the recently studied exotic branes (see \cite{deBoer:2012ma} and references therein).

This reformulation of ten-dimensional supergravity is based on a field redefinition, so we believe that the symmetries should be preserved in the reformulation, even if possibly appearing in a more complicated fashion. This is also an interesting point to study. The resulting new supergravity would provide an uplift to some four-dimensional gauged supergravities, that did not have any before. It would then be nice to find new interesting solutions directly from ten dimensions, in particular de Sitter solutions. Due to the non-triviality of the global aspects, some new physics could be accessed that way. Indeed, suppose we have a solution of $\tL$, which satisfies the usual compactification ansatz. It can be that its expression in terms of the standard NSNS fields is too complicated to have been considered before (especially because of its global properties, for instance if it is non-geometric). If in addition it is not T-dual to any geometric solution, this solution can fairly be thought of as new. As a solution of $\tL$, it then provides new interesting physics (see addendum 3). We hope to come back to these ideas in future work.\\

\vspace{0.3in}

\noindent {\bf Addendum 1}: The reformulation of the standard ten-dimensional NSNS Lagrangian thanks to the field redefinition, discussed in sections \ref{sec:fieldredef} and \ref{sec:rewrit}, was clarified in \cite{Andriot:2013xca}. There, the resulting Lagrangian $\tL_{\b}$ (it differs from $\tL$ here by a total derivative) was said to correspond to the NSNS sector of a theory called for convenience $\beta$-supergravity. The form of $\tL_{\b}$ was simplified with respect to $\tL$ thanks to a better identification of the fluxes: those are better defined in flat (tangent space) indices as
\beq
Q_{c}{}^{ab} = \del_c \b^{ab} - 2 \b^{d[a} f^{b]}{}_{cd}\ ,\quad R^{abc} = 3 \b^{d[a}\nabla_d \b^{bc]} \ , \quad f^{a}{}_{bc}=2 \te^{a}{}_{m} \del_{[b} \te^{m}{}_{c]}\ . \label{fluxesflat}
\eeq
The $R$-flux being a tensor, its formula \eqref{fluxesflat} is the same the one in \eqref{10dnongeoflux} multiplied by vielbeins $\te^a{}_m$ ($\nabla$ is the usual covariant derivative with Levi-Civita connection). The $Q$-flux in \eqref{fluxesflat} is however not a tensor, and is in general not directly related to the one in \eqref{10dnongeoflux}. This difference in the $Q$-flux, suggested several times in the present paper, was the key point to simplify the Lagrangian. The DFT completion of $\beta$-supergravity (in the form of \cite{Andriot:2013xca}) is achieved in \cite{Geissbuhler:2013uka}, building on \cite{Aldazabal:2011nj}, and the fluxes there match \eqref{fluxesflat} upon the strong constraint. This DFT formulation is related to the Generalized Geometry one, derived in \cite{Andriot:2013xca}. In that paper and in \cite{Geissbuhler:2013uka} is also discussed in more details the matching with four-dimensional gauged supergravity and its potential.\\

\noindent {\bf Addendum 2}: The objects presented in section \ref{sec:diffeocov} were further developed in \cite{Andriot:2013xca}, although only at the supergravity level (meaning with $\tilde{\del}^m=0$). In particular, the derivative $\tilde{\nabla}$, denoted there $\widecheck{\nabla}$, was shown to play a crucial role. We considered its flat indices version with a corresponding spin connection $\omega_Q$. Then, the $Q$-flux \eqref{fluxesflat} turned out to play precisely the same role in $\omega_Q$ as the geometric flux $f$ usually plays in the standard spin connection of $\nabla$. So the intuition of the $Q$-flux not being a tensor, but rather playing the role of a connection, here presented for DFT and with a different formula for this flux, actually still held in the subsequent work. The other objects presented here, such as $\widecheck{\R}$, also got further developments and interpretations in \cite{Andriot:2013xca}.\\

\noindent {\bf Addendum 3}: The discussion in sections \ref{sec:global} and \ref{sec:ccl} on global aspects and novelty of the backgrounds described got refined in later work. Because of \eqref{LtL}, the symmetries of $\beta$-supergravity are generically those of standard supergravity, and are therefore restricted for the NSNS sector to diffeomorphisms and $b$-field gauge transformations. As stressed in \cite{Blumenhagen:2013aia}, the symmetries of the theory are in principle the only transformations one should use to patch the fields. Therefore, in general, the description of a non-geometric background would not be improved by going from standard supergravity to $\beta$-supergravity. This statement can however be refined as follows, as pointed out in \cite{Andriot:2013xca, Andriot:2014uda}: if one focuses rather on subcase by making a further assumption or implementing a further constraint, then the symmetries can change and in particular get enhanced. This is precisely what happens here when considering the presence of $n$ isometries: the T-duality group $O(n,n)$ then becomes part of the available symmetries. Of special interest are the subgroup of $\beta$-transforms, under which the Lagrangian of $\beta$-supergravity is manifestly invariant. We proved \cite{Andriot:2014uda} that the whole class of backgrounds of standard supergravity patching with those (together with possible diffeomorphisms) would very likely be non-geometric, while their reformulation in $\beta$-supergravity would be geometric there (in particular the fluxes \eqref{fluxesflat} are left invariant). This class of backgrounds thus realises the scenario described in section \ref{sec:global}, and the toroidal example falls precisely in this class. Considering isometries is thus an important subcase. Unfortunately, we showed as well that the whole class was T-dual to geometric backgrounds of standard supergravity, preventing from getting new physics from them. It is for now unclear whether a different situation could be obtained from $\beta$-supergravity, by for instance considering another subcase, adding other sectors, etc.

% ---------------------------------------- Bibliography

%\newpage

\providecommand{\href}[2]{#2}\begingroup\raggedright
\endgroup

\end{document}